\documentclass[runningheads]{llncs}
\usepackage[T1]{fontenc}
\usepackage{graphicx}
\usepackage{caption}
\usepackage{subcaption}
\usepackage{booktabs}
\usepackage{multirow}
\usepackage{graphicx}
\usepackage{hyperref}
\usepackage[subtle,margins=normal,leading=normal]{savetrees}

\begin{document}
\title{Magnetic Resonance Imaging Feature-Based Subtyping and Model Ensemble for Enhanced Brain Tumor Segmentation}

%
\titlerunning{Enhanced Brain Tumor Segmentation in MRI}
%
\author{ Zhifan Jiang\inst{1}*, Daniel Capell\'{a}n-Mart\'{i}n\inst{1,2}*, Abhijeet Parida\inst{1,2}*,  \\ Austin Tapp\inst{1}, Xinyang Liu\inst{1},  Mar\'{i}a J. Ledesma-Carbayo\inst{2},\\ Syed Muhammad Anwar\inst{1,3}  and Marius George Linguraru\inst{1,3}}

\authorrunning{Z. Jiang, D. Capell\'{a}n-Mart\'{i}n,  A. Parida et al.}
%
\institute{Children’s National Hospital, Washington, DC, USA \and Universidad Polit\'{e}cnica de Madrid and CIBER-BBN, ISCIII, Madrid, Spain \and George Washington University, Washington, DC, USA}
\maketitle              
\begin{abstract}
Accurate and automatic segmentation of brain tumors in multi-parametric magnetic resonance imaging (mpMRI) is essential for quantitative measurements, which play an increasingly important role in clinical diagnosis and prognosis. The International Brain Tumor Segmentation (BraTS) Challenge 2024 offers a unique benchmarking opportunity, including various types of brain tumors in both adult and pediatric populations, such as pediatric brain tumors (PED), meningiomas (MEN-RT) and brain metastases (MET), among others. Compared to previous editions, BraTS 2024 has implemented changes to substantially increase clinical relevance, such as refined tumor regions for evaluation. We propose a deep learning-based ensemble approach that integrates state-of-the-art segmentation models. Additionally, we introduce innovative, adaptive pre- and post-processing techniques that employ MRI-based radiomic analyses to differentiate tumor subtypes. Given the heterogeneous nature of the tumors present in the BraTS datasets, this approach enhances the precision and generalizability of segmentation models. On the final testing sets, our method achieved mean lesion-wise Dice similarity coefficients of 0.926, 0.801, and 0.688 for the whole tumor in PED, MEN-RT, and MET, respectively. These results demonstrate the effectiveness of our approach in improving segmentation performance and generalizability for various brain tumor types. \\

* These authors contributed equally.
\keywords{Brain tumor segmentation \and MRI
\and Deep learning \and Pediatric brain tumors \and Meningiomas \and Metastases.
}
\end{abstract}

\section{Introduction}
Currently, brain cancer is the 10th leading cause of cancer death across all age groups for both males and females in the United States, with an estimated 18,760 deaths from malignant brain tumors \cite{cancer-stats-2024}. Early and accurate diagnosis is crucial for effective treatment planning, which can significantly impact patient outcomes. The variability in tumor appearance across different imaging modalities further complicates assessing and managing these conditions, underscoring the critical need for precise and reliable diagnostic tools.

In this context, segmentation of brain tumors using deep learning techniques has emerged as a transformative approach. Traditional manual segmentation methods are labor-intensive, time-consuming, and subject to inter-observer variability, leading to inconsistencies in clinical decision-making. Deep learning models, particularly those leveraging convolutional neural networks (CNNs), offer a robust and automated solution for brain tumor segmentation by learning complex patterns from large datasets \cite{bakas2018}. These models can achieve high accuracy and consistency, facilitating early detection, precise localization, and detailed characterization of tumors. By incorporating various MRI-based scanning protocols, deep learning models can successfully account for diverse morphological and pathological brain tumor features thereby enhancing diagnostic accuracy and improving treatment planning \cite{zhou2020}. Ultimately,  deep learning for automatic brain tumor segmentation promises to improve patient outcomes by enabling more personalized and data-driven therapeutic interventions \cite{kamnitsas2017,zhou2020}.

This paper presents a methodology developed primarily for the segmentation of pediatric brain tumors (PED); the method was adapted for the segmentation of meningioma radiotherapy (MEN-RT) and brain metastases (MET). 

\section{Segmentation Challenges}
 Since 2012, the international brain tumor segmentation (BraTS) challenge~\cite{brats2021,bakas1,bakas2,bakas3,medperf,brats2015}, held in conjunction with the international conference on Medical Image Computing and Computer Assisted Intervention (MICCAI), has generated a benchmark dataset for the segmentation of adult brain gliomas. The BraTS 2024~\cite{brats2021} is expanded to a cluster of challenges, encompassing a variety of tumor types alongside histopathology and augmentation tasks. Herein, we propose a generalized segmentation technique for PED, MEN-RT, and MET.

\noindent\textbf{PED:}
The BraTS-PEDs 2024 challenge \cite{kazerooni2024braintumorsegmentationpediatrics,PEDarxiv} aims to address the unique challenges of pediatric brain tumors, including diffuse midline gliomas and high-grade astrocytoma. The data (N=464) were collected from institutions including the Children's Brain Tumor Network (CBTN, N=120), DMG/DIPG Registry (N=256), Boston's Children Hospital (N=61), and Yale University (N=27). 

Currently, this challenge consists of 261 training, 91 validation, and 86 testing cases. Each case includes pre-contrast T1-weighted (T1), constrast-enhanced T1-weighted (T1CE), T2-weighted (T2), and T2-weighted fluid-attenuated inversion recovery (T2-FLAIR) MRI sequences. The task focuses on the segmentation of the whole tumor (WT), tumor core (TC), enhancing tumor (ET), cystic components (CC), non-enhancing tumor core (NET), and edema (ED).

\noindent\textbf{MEN-RT:}
The BraTS 2024 MEN-RT Challenge~\cite{MENarxiv,MENarxiv2024} aims to advance automated segmentation of the gross tumor volume (GTV) for meningiomas using pre-radiation therapy MRI scans. Meningiomas, the most common primary intracranial tumors, vary in grade and higher grades are associated with greater morbidity and frequent recurrence~\cite{MENarxiv}. Accurate GTV segmentation is crucial for effective radiation therapy but remains complex and necessitates intensive expert delineation. 

Unlike other sub-challenges that focus on pre-operative cases, this challenge uses one 3D T1CE MRI sequence in native acquisition space. Faces are automatically removed for patient anonymity and the skull is left skull intact, unlike traditional BraTS data that is typically skull-stripped~\cite{MENarxiv}. This challenge consists of 500 training, 70 validation, and 180 testing cases focusing on benchmark algorithms for precise GTV delineation, aiding radiotherapy planning and supporting future research on tumor progression and recurrence.

\noindent\textbf{MET:}
The BraTS 2024 MET challenge \cite{METarxiv} addresses the labor-intensive and error-prone process of monitoring metastatic brain disease, particularly when lesions are small and hard to segment. Traditional methods often focus on measuring the largest diameter of metastases, but accurate volumetric estimates are crucial for clinical decision-making and treatment prediction. 

This challenge aims to benchmark algorithms for automatically segmenting brain metastases (WT, TC, and ET) and surrounding ED, improving efficiency and consistency. The dataset consists of 651 training, 88 validation, and 119 testing cases with T1, T1CE, T2, and T2-FLAIR sequences.

\section{Methods}
\begin{figure}[htbp]
    \centering
    \includegraphics[width=\linewidth]{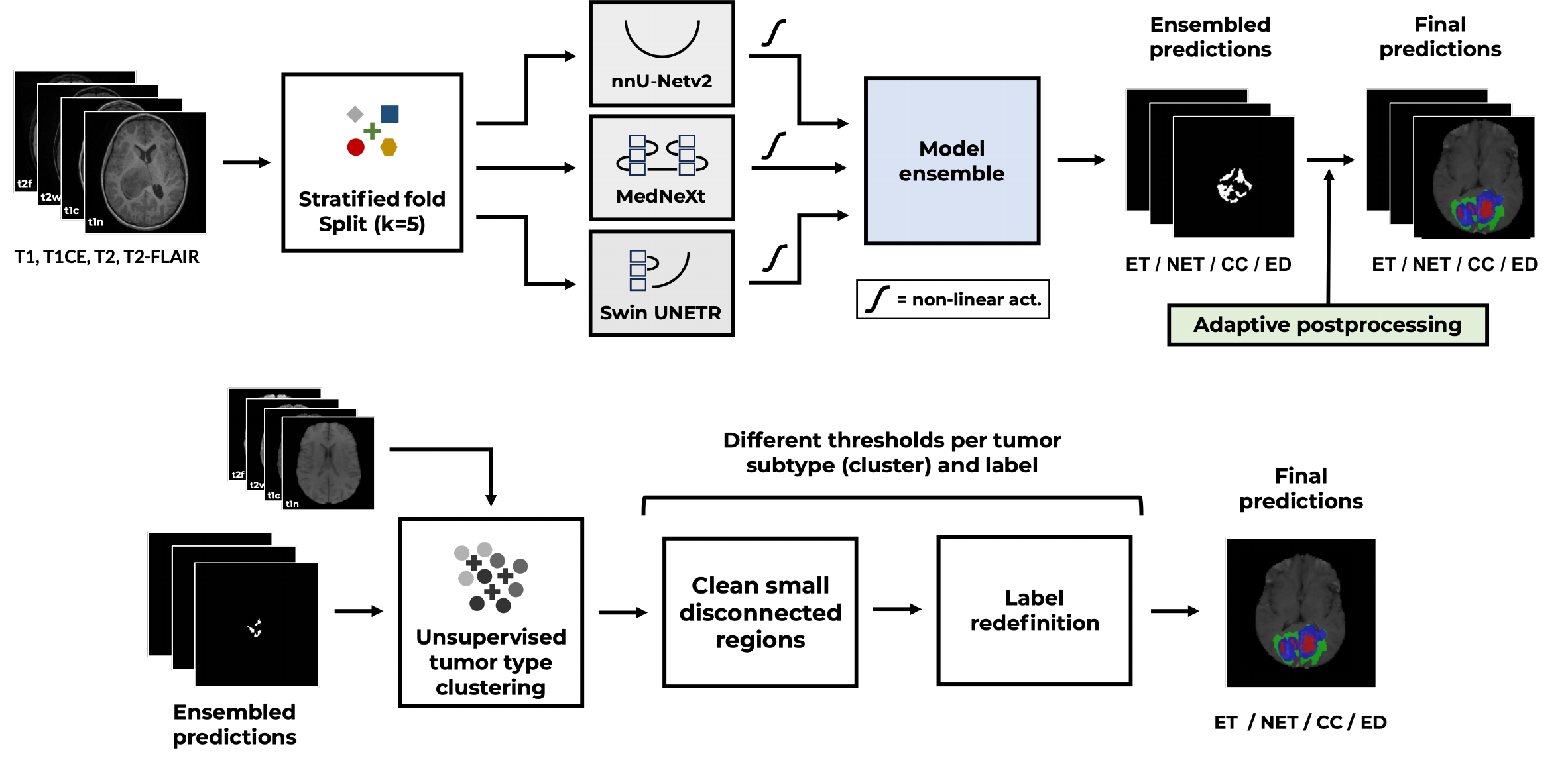}
    \caption{\textbf{Proposed Method}: Unsupervised stratified fold splitting, model training, model ensemble, and adaptive post-processing. Outputs are generated using three state-of-the-art deep learning models, which are processed through ensemble strategies. The ensembled predictions are then refined using a specifically tailored adaptive post-processing step.}
    \label{fig:pipeline}
\end{figure}

Our methodology utilizes a common segmentation framework, wherein each component is tailored to specific BraTS challenges, which present different types of brain tumors. Consequently, the methodology is generalizable and can be easily adapted to new types of brain tumors to optimize their segmentation. The framework (Figure \ref{fig:pipeline}) includes data pre-processing, model training, model ensembling, and post-processing of predicted labels. It also incorporates MRI radiomic feature-based subtype clustering for both training and post-processing.

\subsection{MRI Radiomic Features for Tumor Subtypes}
\label{subsec:radiomics}
We defined tumor subtypes for each task using MRI radiomic features on the segmented tumor regions of interest in each input MRI sequence. We utilized the PyRadiomics package~\cite{van_griethuysen_computational_2017} and its implementation in \cite{jiang_automatic_2023} to extract these features. Specifically, radiomic features were computed on the largest lesion area (connected component of the WT) for each case. These features were categorized into two groups: 14 shape- and 93 intensity-based features per MRI sequence. Since some radiomic features are sensitive to image spatial resolution, input images were resampled to an isotropic spacing (0.9375 mm for MEN-RT and 1.0 mm for MET, respectively) before applying radiomic measurements. 

Principal component analysis was used to select the most relevant features, explaining 99\% of the variance, resulting in 9 features for PED, 3 for MEN-RT, and 2 for MET. The k-means clustering algorithm grouped lesions into different clusters (subtypes) of tumors based on the most relevant radiomic features. An optimal number of clusters was determined using grid search and silhouette analysis on the k-means clustering results. This k-means algorithm was trained on the training set with their corresponding ground-truth WT for tumor subtype analysis during pre-processing, as well as the cross-validated predicted WT during post-processing.

\subsection{Model Training}
\subsubsection{nnUNet} is based on the U-Net architecture \cite{unet} and is a self-configuring deep learning framework for semantic segmentation. According to the specific imaging modality and unique attributes of each dataset, the framework autonomously adjusts its internal configurations, resulting in an improved segmentation performance and generalization when compared to other state-of-the-art methods for biomedical image segmentation \cite{nnunet}.

We trained a full-resolution 3D nnU-Net (v2) model on the stratified five-fold created for the tasks of PED, MEN-RT, and MET. A preprocessing of zero mean unit variance normalization was applied to the input images. Each input image was divided into patches of $128\times128\times128$ voxels. For each task, PED, MEN-RT, or MET, we trained models to predict the respective number of task-dependent output labels, i.e., 4 labels were predicted for PED, 1 label was predicted for MEN-RT, and 3 labels were predicted for MET.

We used a class-weighted loss function combining Dice loss and cross-entropy loss. To optimize this loss function, we used the stochastic gradient descent (SGD) optimizer with Nesterov momentum, using an initial learning rate of 0.01, momentum of 0.99, and weight decay of 3e-05. Each of the five folds was trained for 200 epochs on NVIDIA A100 (40 GB) and NVIDIA V100 (16GB) GPUs. During inference, images were predicted using a sliding window approach, with the window size matching the patch size used during training. 
The nnU-Net implementation is available in an open-source repository: \href{https://github.com/MIC-DKFZ/nnUNet}{https://github.com/MIC-DKFZ/nnUNet}.

\subsubsection{MedNeXt}\cite{mednext} leverages a hybrid approach combining convolutional neural networks and attention mechanisms for medical image analysis. The architecture integrates convolutional layers for feature extraction with attention modules that enhance the model’s focus on relevant regions within the images \cite{mednext}. Based on  3D nnU-Net (v2)\cite{nnunet} strategies the framework autonomously adjusts its internal configurations to give better performances.

As with 3D nnU-Net, MedNeXt was trained in a label-respective manner for each task. For all tasks, we trained MedNeXt-M (k=3, 17.6M parameters, 248 GFlops) with a class-weighted loss function combining Dice loss and cross-entropy loss. To optimize this loss function, we used the stochastic gradient descent (SGD) optimizer with Nesterov momentum, using an initial learning rate of 0.01, momentum of 0.99, and weight decay of 3e-05. Each of the five folds was trained for 200 epochs on NVIDIA A100 (40 GB) and NVIDIA V100 (16GB) GPUs. During inference, images were predicted using a sliding window approach, with the window size matching the patch size used during training. The MedNeXt implementation is available in an open-source repository: \href{https://github.com/MIC-DKFZ/MedNeXt}{https://github.com/MIC-DKFZ/MedNeXt}.

\subsubsection{SwinUNETR} is a vision transformer-based \cite{vit} hierarchical structure for localized self-attention using shifted windows \cite{swinunetr2,unetr,swinunetr}. We trained a 3D SwinUNETR model using five-fold cross-validation for each task. 

Each input image was sampled four times using patches of $128\times128\times128$ voxels and batch size of 1 was used to fully utilize the GPU’s memory. The model output was $n+1$ channels corresponding to the $n$ non-overlapping labels and background. Softmax activations were used at the output layer. We used a class-weight loss function that combined Dice loss and focal loss. To optimize the loss function, we used the AdamW optimizer with an initial learning rate of 0.0001, momentum of 0.99, and weight decay of 3e-05. Each of the folds was trained on an NVIDIA H100 (80 GB) GPU and an NVIDIA A6000 (48GB) GPU. The number of epochs varies across tasks: 650 epochs for PED, 250 epochs for MEN-RT and MET. The Swin UNETR implementation is part of the PyTorch-based framework MONAI: 
\href{https://monai.io}{https://monai.io}.

\subsection{Model Ensemble}
We used a model ensemble strategy to enhance the accuracy and robustness of the segmentation outcome \cite{brats2023sub}. This approach involves harnessing the complementary strengths of the models described, nnU-Net, MedNeXt, and SwinUNETR, to improve the probability prediction for the segmentation task.
\begin{equation}
    Y_{ensemble} = w_n * nnUNet(X) + w_m * MedNeXt(X) + w_s * SwinUNETR(X)
\end{equation}

The weights $w_n$ for nnUNet,  $w_m$ for MedNeXt, and $w_s$ for SwinUNETR are estimated using the individual model performance on the training set using five-fold cross validation. For the PED, we obtained $w_n=0.33$, $w_m=0.34$, and $w_s=0.33$. While for the MEN-RT, we obtained $w_n=0.33$,  $w_m=0.33$, and $w_s=0.34$. Finally, for the MET, given the significantly lower  performance of SwinUNETR, we employed $w_n=0.487$, $w_m=0.513$, and $w_s=0$.

\subsection{Post-processing}
Following the strategy outlined in \cite{brats-goat}, we utilized an adaptive post-processing technique on the cross-validated WT predictions computed by the ensemble of the models trained on each fold. 
After the k-means algorithm was trained on the training set images (Section \ref{subsec:radiomics}), we performed an optimal threshold search to eliminate small, disconnected components, thereby reducing the number of false positives in the segmentation maps. These thresholds were determined adaptively within each tumor subtype and for each label separately. Finally, a second threshold search was conducted on these refined segmentation maps to redefine labels, i.e., labels which ratio with respect to WT fell below certain threshold, the label would be redefined to other label.

Details regarding thresholds and weights for model ensemble can be found in the source code of our implementation available at 
\href{https://github.com/Precision-Medical-Imaging-Group/HOPE-Segmenter-Kids}{https://github.com/Precision-Medical-Imaging-Group/HOPE-Segmenter-Kids}. Additionally, an open-source web-based application for demonstration purposes is accessible at \href{https://segmenter.hope4kids.io/}{https://segmenter.hope4kids.io/}, allowing users to obtain segmentation and volumetric measurements by uploading MRI sequences.

\section{Results}
\subsection{Metrics}
The evaluation of the model prediction on the validation set was done on the BraTS pipeline on the Synapse platform. The models were assessed for each of the regions using the lesion-wise Dice score and the $95^{th}$ percentile lesion-wise Hausdorff distance. 

\subsection{Segmentation Performance}

\begin{figure}[h!]
    \centering
  \includegraphics[width=1.0\textwidth]{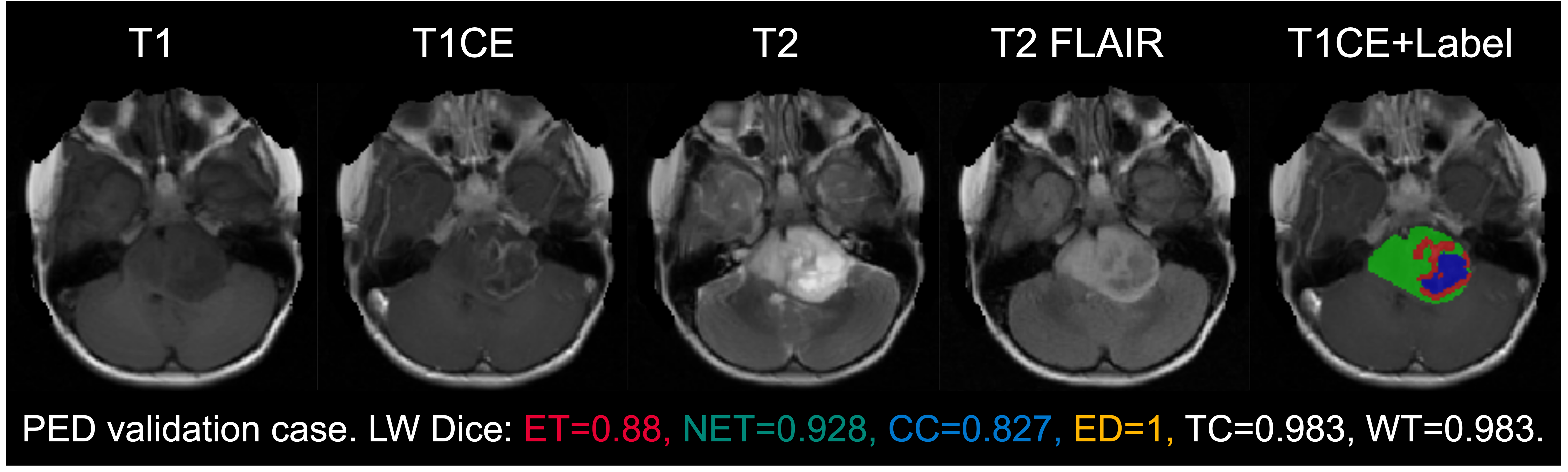}
  \caption{\textbf{Qualitative Results}: The model ensemble, after post-processing, was evaluated on a PED validation case (BraTS-PED-00300-000). Numbers represent the segmentation performance in lesion-wise (LW) Dice for different tumor regions: enhancing tumor (ET) in red, non-enhancing tumor (NET) in green, cystic components (CC) in blue, edema (ED) in yellow, combined tumor core (TC), and whole tumor (WT).}
  \label{fig:qual-results-ped}
\end{figure}

\begin{figure}[h!]
    \centering
  \includegraphics[width=0.45\textwidth]{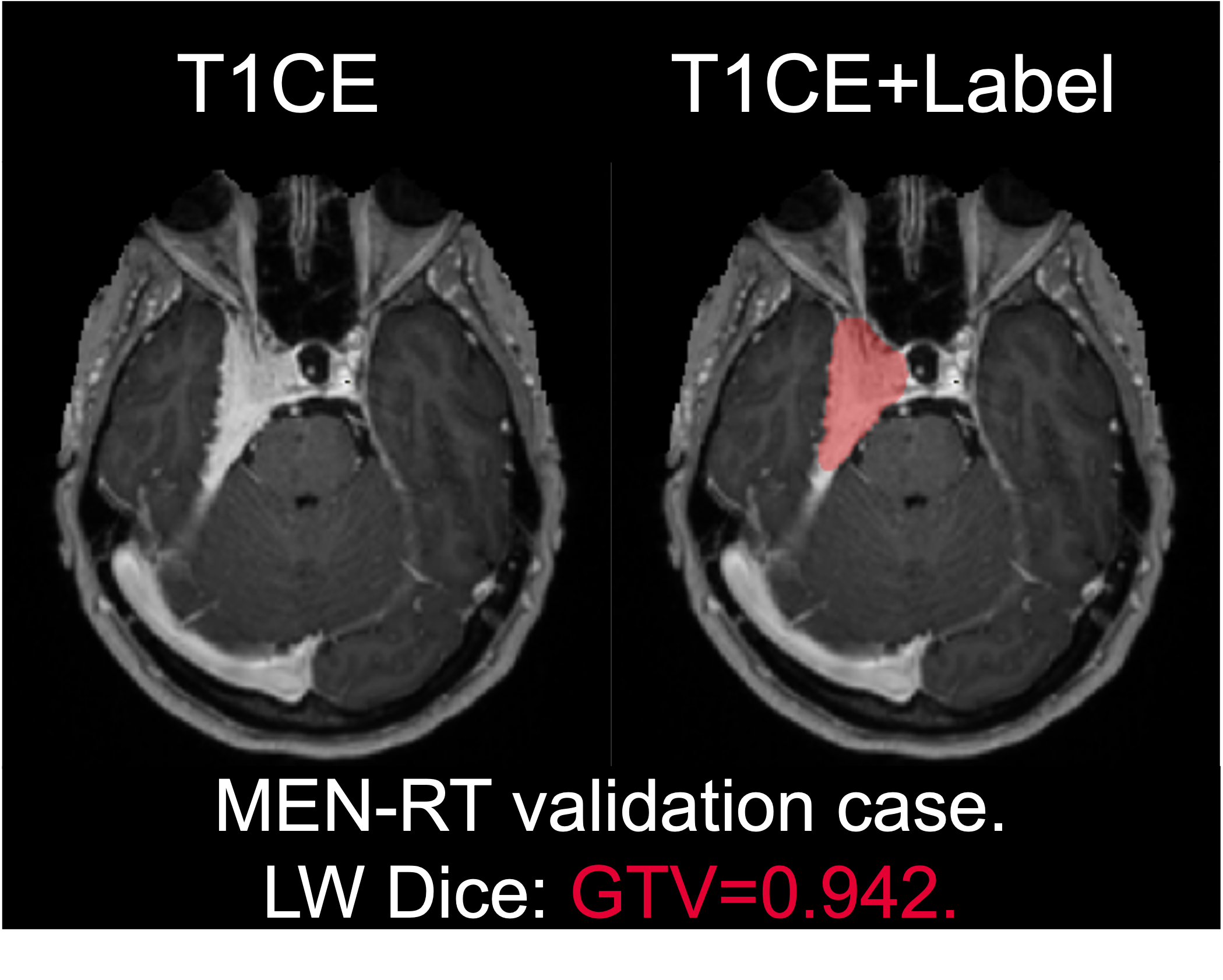}
  \caption{\textbf{Qualitative Results}: The model ensemble, after post-processing, was evaluated on an MEN-RT validation case (BraTS-MEN-RT-0698-1). The number represents the segmentation performance in lesion-wise (LW) Dice for the gross tumor volume (GTV) in red.}
  \label{fig:qual-results-men}
\end{figure}

\begin{figure}[h!]
    \centering
  \includegraphics[width=1.0\textwidth]{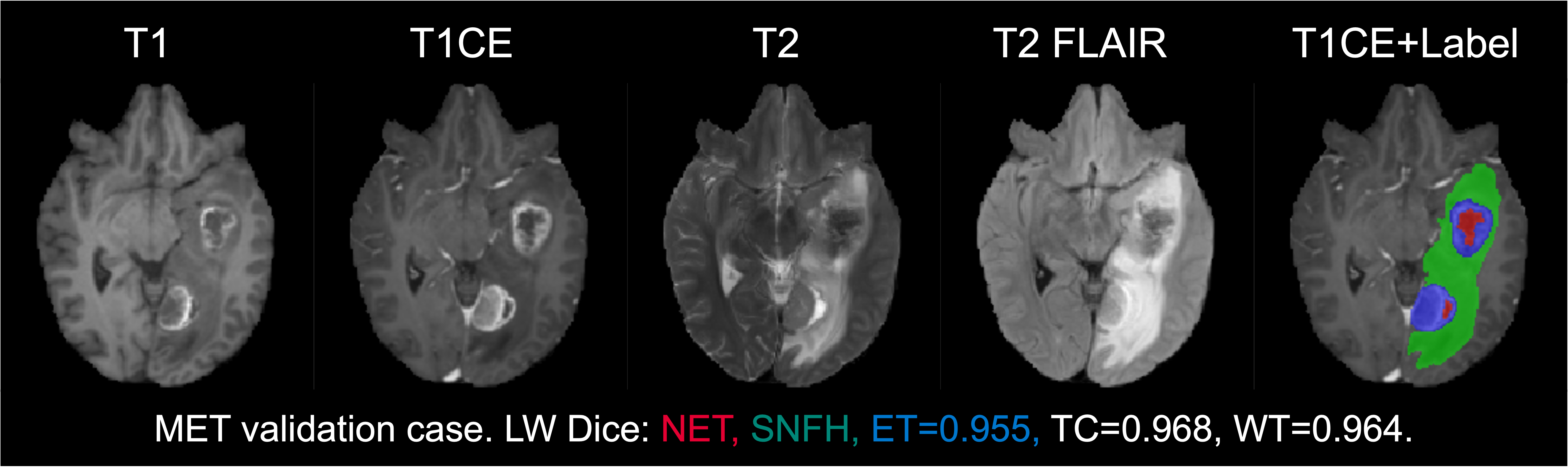}
  \caption{\textbf{Qualitative results}: The model ensemble, after post-processing, was evaluated on an MET validation case (BraTS-MET-00908-000). Numbers represent the segmentation performance in lesion-wise (LW) Dice for different tumor regions: enhancing tumor (ET) in blue, combined tumor core (TC), and whole tumor (WT). Labels represent non-enhancing tumor (NET) in red and surrounding non-enhancing FLAIR hyperintensity (SNFH) in green.}
  \label{fig:qual-results-met}
\end{figure}

Quantitative results of our models across the validation and testing datasets for each challenge are shown in Tables \ref{tab:val-results-peds}, \ref{tab:val-results-men}, and  \ref{tab:val-results-met} for PED, MEN-RT, and MET, respectively. These evaluations were performed automatically by the challenge's digital platform, with no access to the validation ground truth data and no access to any testing data including images and labels. Additionally, Figures~\ref{fig:qual-results-ped}, ~\ref{fig:qual-results-men}, and~\ref{fig:qual-results-met} illustrated qualitative results on validation cases for PED, MEN-RT, and MET, respectively.

\begin{table}[h!]
\caption{\textbf{Quantitative results} on the validation and testing datasets of PED. Lesion-wise (LW) Dice coefficients and 95\% Hausdorff distances (HD95) were computed for enhancing tumor (ET), tumor core (TC), whole tumor (WT), non-enhancing tumor (NET), cystic components(CC), and edema (ED), respectively.}
\centering
\resizebox{1.0\textwidth}{!}{%
\begin{tabular}{@{}clllccccccclcccccc@{}}
\toprule
\multirow{2}{*}{\textbf{Task}} &  & \multicolumn{1}{l}{\multirow{2}{*}{\textbf{Model}}} & \textbf{} & \multicolumn{6}{c}{\textbf{LW Dice}} & \textbf{\hspace{0.2cm}} & \multicolumn{6}{c}{\textbf{LW HD95 (mm)}} \\ \cmidrule(l){4-17} 
 &  & \multicolumn{1}{c}{} & \textbf{} & \textbf{ET} & \textbf{TC} & \textbf{WT} & \textbf{NET}& \textbf{CC}& \textbf{ED}&\textbf{} & \textbf{ET} & \textbf{TC} & \textbf{WT} & \textbf{NET}& \textbf{CC}& \textbf{ED} \\ \midrule
&  & MedNeXt &  & 0.68 &0.931&0.931&0.903&0.73&0.879&& 70.04&11.01&10.97&11.67&94.88&45.21 \\
\textbf{PED} &  & nnU-Net &  &  0.608&0.93&0.93&0.9&0.742&0.846&&98.88&8.91&8.88&9.74&67.49&57.54\\
\textbf{Validation} &  & SwinUNETR &  & 0.607 & 0.921 & 0.921&0.892&0.683&0.967 & &91.02 &  9.18 & 9.18 &10.2& 115.24&12.33 \\
 \textbf{N=91} &  & Ensemble &  & 0.68& 0.931 & 0.931&0.904&0.697&0.956	&&67.93&8.93&8.91&9.79&103.17&16.44  \\
&  & Post-processing &  & {0.665} & {0.931} & {0.931} & {0.904} & {0.701} & {0.967} & & {78.11} & {8.93} & {8.93} & {9.8} & {111.06} & {12.33} \\ 
\midrule
&  & Mean &  & 0.692 & 0.918 & 0.926 & 0.859 &0.715& 0.884 && 53.87 & 7.46 & 7.14 & 8.01 & 96.41 & 43.49 \\
\textbf{PED} &  & (Standard deviation) &  & (0.328) & (0.124) & (0.117) & (0.198) & (0.44) & (0.322) && (126.25) & (40.16) & (40.15) & (40.11) & (163.73) & (120.59) \\
\textbf{Testing} &  &25th quantile &  & 0.611 & 0.876 & 0.89 & 0.803 & 0.001 & 1 & & 1 &  1 & 1 & 1 & 0 & 0 \\
 \textbf{N=86} &  & Median &  & 0.816 & 0.965 & 0.967 & 0.947 & 1 & 1	&& 2 & 1.41 & 1.21 & 2.24 & 0 & 0  \\
&  & 75th quantile &  & 0.931 & 0.982 & 0.982 & 0.973 & 1 & 1 & & 8 & 4.12 & 3.94 & 5.17 & 285.41 & 0 \\ 

\bottomrule
\end{tabular}%
}
\label{tab:val-results-peds}
\end{table}

\begin{table}[h!]
\caption{\textbf{Quantitative results} on the validation and testing datasets of MEN-RT. Lesion-wise (LW) Dice coefficients and 95\% Hausdorff distances (HD95) were computed for gross tumor volume (GTV).}
\centering
\begin{tabular}{@{}clllclc@{}}
\toprule
\multirow{1}{*}{\textbf{Task}} &  & \multicolumn{1}{l}{\multirow{1}{*}{\textbf{Model}}} & \textbf{} & \multicolumn{1}{c}{\textbf{LW Dice GTV}} & \textbf{} & \multicolumn{1}{c}{\textbf{LW HD95 (mm) GTV}} \\  \midrule
&  & MedNeXt &  & 0.803 & & 29.68  \\
\textbf{MEN-RT} &  & nnU-Net &  & 0.764	&& 59.41 \\
\textbf{Validation} &  & SwinUNETR &  & 0.761 && 48.87 \\
\textbf{N=70} &  & Ensemble &  & 0.794 & &  43.81 \\
&  & Post-processing &  & 0.794 & &  43.81 \\
\midrule
&  & Mean &  & 0.801 & & 39.24 \\
\textbf{MEN-RT} &  & (Standard deviation) &  & (0.21) && (147.08) \\
\textbf{Testing} &  & 25th quantile &  & 0.753 && 1.02 \\
\textbf{N=180} &  & Median &  & 0.871 & & 2.27 \\
&  & 75th quantile &  & 0.94 & & 6.57 \\
\bottomrule
\end{tabular}

\label{tab:val-results-men}
\end{table}

\begin{table}[h!]
\caption{\textbf{Quantitative results} on the validation and testing datasets of MET. Lesion-wise (LW) Dice coefficients and 95\% Hausdorff distances (HD95) were computed for enhancing tumor (ET), tumor core (TC), and whole tumor (WT), respectively.  *Due to the significantly lower performance of SwinUNETR, we included MedNeXt and nnU-Net for model ensemble.}

\centering
\begin{tabular}{clllccclccc}
\toprule
\multirow{2}{*}{\textbf{Task}} &  & \multicolumn{1}{l}{\multirow{2}{*}{\textbf{Model}}} & \textbf{} & \multicolumn{3}{c}{\textbf{LW Dice}} & \textbf{} & \multicolumn{3}{c}{\textbf{LW HD95 (mm)}} \\ \cmidrule(l){4-11} 
 &  & \multicolumn{1}{c}{} & \textbf{} & \textbf{ET} & \textbf{TC} & \textbf{WT} & \textbf{} & \textbf{ET} & \textbf{TC} & \textbf{WT} \\ \midrule
&  & MedNeXt &  & 0.746 & 0.762 & 0.725 & & 52.11 &	51.99 &	53.11 \\
\textbf{MET} &  & nnU-Net &  & 0.735 & 0.753 & 0.707 & & 54.44 & 54.28 &	60.99 \\
\textbf{Validation} &  & SwinUNETR&  & 0.646 & 0.667 & 0.618 & & 83.8 & 83.7 & 93.7 \\
 \textbf{N=88} &  & Ensemble* &  & 0.739 & 0.756 & 0.717 & & 52.37 & 52.22 & 55.61 \\
 &  & Post-processing* &  & 0.739 & 0.755 & 0.719 & & 52.37 & 52.23 & 53.53 \\
 \midrule
&  & Mean &  & 0.675 & 0.697 & 0.688 &	& 68.39 & 68.21 & 64.88 \\
\textbf{MET} &  & (Standard deviation) &  & (0.225) & (0.23) & (0.247) &	& (89.12) & (89.18) &	(88.95) \\
\textbf{Testing} &  & 25th quantile &  & 0.499 & 0.506 & 0.487 & & 1.17 & 1 & 1.41 \\
\textbf{N=119} &  & Median &  & 0.701 & 0.714 & 0.743 & & 3 & 2.83 & 3.71 \\
 &  & 75th quantile &  & 0.881 & 0.915 & 0.919 & & 126.07 & 126.03 & 126.27 \\
 \bottomrule
\end{tabular}

\label{tab:val-results-met}
\end{table}

\section{Discussion}
Our observations indicated that ensemble techniques generally enhance the precision and generalizability of segmentation models. Specifically, we found that weighted ensembles outperformed simple averaged ensembles of probabilities. By assigning different weights to the outputs of individual models based on their validation performance, the weighted ensemble method effectively prioritizes more accurate predictions, leading to improved overall segmentation results. However, the improvement is limited in some tasks.

Despite these improvements, the variability of tumor regions across subjects remains a significant challenge. The heterogeneous nature of brain tumors, characterized by differences in size, shape, and location, complicates the segmentation task. This variability suggests that a one-size-fits-all post-processing approach may not be optimal.

To address this issue, we propose incorporating cluster-wise post-processing techniques. By clustering similar tumor subtypes based on radiomic features, we can tailor the post-processing steps to the specific characteristics of each cluster. This approach allows for more precise adjustments to the segmentation results, potentially reducing false positives and improving the overall accuracy of tumor segmentation.

\section{Conclusion}
This year's challenge introduced changes to enhance clinical relevance, such as refined tumor region evaluations. We developed a deep learning-based ensemble method that integrates advanced segmentation models, complemented by an innovative adaptive pre- and post-processing technique utilizing MRI-based radiomic analysis to distinguish tumor subtypes. This approach addresses the heterogeneous nature of tumors in clinical datasets, improving the precision and generalizability of segmentation models. Our method is ready to be tested on other clinical datasets to investigate the efficacy and robustness of our approach in enhancing segmentation accuracy and generalizability across various brain tumor types.



\begin{credits}
\subsubsection{\ackname} Partial support for this work was provided by the National Cancer Institute (UG3 CA236536) and by the Spanish  Ministerio de Ciencia e Innovación, the Agencia Estatal de Investigación, NextGenerationEU funds, under grants PDC2022-133865-I00 and PID2022-141493OB-I00, and EUCAIM project co-funded by the European Union (Grant Agreement \#101100633). The authors gratefully acknowledge the Universidad Politécnica de Madrid (www.upm.es) for providing computing resources on the Magerit Supercomputer.

\subsubsection{\discintname}
The authors have no competing interests to declare that are
relevant to the content of this article.
\end{credits}

%
%
%
\bibliographystyle{splncs04}
\bibliography{references}

\end{document}